\documentclass[conference]{IEEEtran}
\IEEEoverridecommandlockouts
\usepackage{cite}
\usepackage{amsmath,amssymb,amsfonts}
\usepackage{algorithmic}
\usepackage{longtable}
\usepackage[linesnumbered,ruled,vlined]{algorithm2e}
\usepackage{graphicx}
\usepackage{textcomp}
\usepackage{xcolor}
\usepackage{lipsum}
\usepackage{amsthm}
\usepackage{makecell}
\usepackage[tableposition=top]{caption}

\usepackage[labelfont=bf]{caption}
\usepackage{enumitem}
\usepackage{subcaption}
\newtheorem{theorem}{Theorem}
\newtheorem{lemma}{Lemma}

\SetKwInput{KwInput}{Input}
\SetKwInput{KwOutput}{Output}
\newtheorem{definition}{Definition}
\def\BibTeX{{\rm B\kern-.05em{\sc i\kern-.025em b}\kern-.08em
    T\kern-.1667em\lower.7ex\hbox{E}\kern-.125emX}}

\long\def \ignoreme#1{}

\newcommand{\tk}[1]{{\bf {\textcolor{blue}{#1 -- TK}}}}

\newcommand{\hoang}[2]{{\bf {\textcolor{olive}{#1 -- HN}}}}
\newcommand{\mt}[1]{{\textcolor{purple}{#1 -- MT}}}

\begin{document}
\title{ATOM: An Efficient Topology Adaptive Algorithm for Minor Embedding in Quantum Computing\\
}

\author{\IEEEauthorblockN{Hoang M. Ngo}
\IEEEauthorblockA{
\textit{University of Florida}\\
hoang.ngo@ufl.edu}
\and
\IEEEauthorblockN{Tamer Kahveci}
\IEEEauthorblockA{
\textit{University of Florida}\\
tkahveci@ufl.edu}
\and
\IEEEauthorblockN{My T. Thai}
\IEEEauthorblockA{
\textit{\textit{University of Florida}}\\
mythai@cise.ufl.edu}
}

\maketitle



\begin{abstract}
Quantum annealing (QA) has emerged as a powerful technique to solve optimization problems by taking advantages of quantum physics. In QA process, a bottleneck that may prevent QA to scale up is minor embedding step in which we embed optimization problems represented by a graph, called logical graph, to Quantum Processing Unit (QPU) topology of quantum computers, represented by another graph, call hardware graph. Existing methods for minor embedding require a significant amount of running time in a large-scale graph embedding. 
To overcome this problem, in this paper, we introduce a novel notion of adaptive topology which is an expandable subgraph of the hardware graph. From that, we develop a minor embedding algorithm, namely Adaptive TOpology eMbedding (ATOM). ATOM iteratively selects a node from the logical graph, and embeds it to the adaptive topology of the hardware graph. Our experimental results show that ATOM is able to provide a feasible embedding in much smaller running time than that of the state-of-the-art without compromising the quality of resulting embedding.~\footnote{This work is partially funded by NSF under Award 2111679 and 1908594.}
\end{abstract}
\begin{IEEEkeywords}
quantum annealing, adaptive hardware topology
\end{IEEEkeywords}

\vspace*{-.5cm}
\section{Introduction}\label{sec:intro}

\ignoreme{
\mt{H, due to the space limit, please re-write this introduction in this way:}

1st paragraph: Directly talking about QA and is benefit. Dont bother to mention gatebase. Quantum computing in general, QA in particular, has shown benefit...

2nd paragraph: Introduce minor embedding, what is this, and why it's so important to speed up the time etc..

3rd paragraph: limitation of existing work. And challenges

4th paragraph: our solution and contribution.

5th paragraph: Organization of the paper.
}

Quantum computing has shown its supremacy over classical computers as it exponentially reduces the running time of certain computational tasks \cite{Arute2019} in cybersecurity \cite{Ahmed2021}, machine learning\cite{Jerbi2021}, and optimization\cite{Guillaume2022} among others. One of the most outstanding paradigms for quantum computing is Quantum Annealing (QA). It focuses solely on solving optimization problems by utilizing quantum fluctuation effect. QA scales to significantly larger number of qubits. This characteristic enables QA to solve large optimization tasks such as scheduling deep space network \cite{Guillaume2022}, designing logistic network \cite{Ding2021}, finding optimal traffic flow \cite{Neukart2017}, and TDMA scheduling in wireless sensor networks \cite{Ishizaki2019}, to name a few. 

QA-inspired quantum computers solve an optimization problem in three steps. First, the quadratic unconstrained binary optimization (QUBO) representation of the given problem is encoded by a graph, called \emph{logical graph}. Second, the resulting logical graph is embedded into Quantum Processing Unit (QPU) whose topology is represented by another graph, called \emph{hardware graph} such that the logical graph can be obtained from the embedded subgraph of the hardware graph by edge contraction. This embedding process is called \textbf{minor embedding}. 
Finally, quantum computers run the quantum annealing process repeatedly on the embedded QPU to obtain the optimal solution for the given optimization problem. Along with the rapid development in the QPU size, QA-inspired quantum computers are able to solve large optimization problems. The fundamental challenge here is that solving minor embedding problem is non-trivial for the topology of the hardware graph or its induced sub-graphs may not be identical to that of the logical graph. Therefore, embedding process remains to be a bottleneck, especially, when the logical graph size is large. There is an urgent need for an effective minor embedding method which scales to large logical graphs.  

{\bf Related Work.} 
There are two main approaches to solve minor embedding, namely top-down and bottom-up. Top-down approach aims to find embeddings of complete graphs \cite{Choi2011}, \cite{Klymko2012}, \cite{Boothby2016} in the hardware graph. Although the embedding of a complete graph can be the solution for any incomplete graph with smaller or same size, embedding for incomplete graphs, especially sparse graphs may require much fewer qubits than embedding a complete graph. Thus, top-down approach is not effective for embedding sparse logical graphs. To address this problem, Goodrich et al. proposed a post-process the resulting embedding to reduce the number of qubits required \cite{Goodrich2018}. They introduced a new notion of biclique virtual hardware framework, which is constructed by compressing the hardware graph. They find a virtual embedding of the logical graph in the virtual hardware framework using the Odd Cycle Traversal (OCT)-based algorithm. OCT-based method is the-state-of-the-art embedding for general logical graphs. However, the post-process in OCT-based method is not efficient for large logical graphs due to its high complexity.

In contrast, bottom-up approach directly constructs solutions based on the topology of the logical and the hardware graphs. It computes minor embedding either by using Integer Programming (IP) which find an exact solution from predefined constraints \cite{Bernal2020}, or by using a progressive heuristic method which embeds one node of the logical graph to the hardware graph at a time\cite{Cai2014}, \cite{Pinilla2019}. Notably, Minorminer method \cite{Cai2014} commercially developed by D-Wave Inc. scales well for sparse logical graphs. Intuitively, Minorminer quickly finds an infeasible embedding in the hardware graph, and then re-embeds overlapping nodes until it obtains a feasible one. The first disadvantage of Minorminer is that the complexity of their algorithm depends on the input hardware graph size, so the running time grows considerably for inappropriate input hardware graphs. In addition, the number of re-embedding may be large, which can contribute significantly in total running time. 

\textbf{Contributions.} In this paper, we introduce a novel idea of adaptive topology for the hardware graph whose size grows along with the size of embedding as needed. We develop an efficient minor embedding method, called Adaptive TOpology eMbedding (ATOM). Our method follows the bottom-up approach in which we embed one node in the logical graph at a time to an adaptive hardware graph. Unlike existing approaches, ATOM does not require a fixed hardware graph size. This leads to a significant reduction in the running time, and overcomes the aforementioned limitations. 

In our experiments, we compare our method with OCT-based \cite{Goodrich2018}, and Minorminer \cite{Cai2014} which are two state-of-the-art embedding methods in each category as discussed in the related work. 
Our results demonstrate that ATOM requires significant less running time, while the hardware size and the number of qubits required for embedding are comparable to the best settings of hardware size for two other methods.

\textbf{Organization.} The rest of the paper is structured as follows. Section \ref{sec:prelim} introduces the definition of minor embedding and related notions. 
 Our solution, ATOM,  and its theoretical analysis are described in Section \ref{sec:method}. Section \ref{sec:exp} presents our experimental results. Finally, Section \ref{sec:con} concludes the paper.


\section{Preliminaries}\label{sec:prelim}

Here, we first present the fundamental terminology needed to understand how QA solves optimization problems. Next, we describe the minor embedding problem in QA.

Recall from Section \ref{sec:intro} that QA consists of three steps. Below, we elaborate on this process. In the first step of QA, given an optimization problem with the set of binary variables $\mathbf{x} = \{x_0, x_1, ... , x_N\}$ and quadratic coefficients $Q_{i,j}$, we represent the problem with QUBO form, and express the optimal solution for the given problem corresponding to the state with lowest energy of final Hamiltonian as $\mathbf{x}^{\ast}$ such that: 
\ignoreme{
\tk{isn't $i < j$ in the summation below? otherwise $x_1x_2$ and $x_2x_1$ will appear both}
\hoang{I think it is $i < j$, but in papers I read, they write this equation. Let me check again. I think both of them are the same because $Q$ is symmetric.}
}
\begin{equation}
    \mathbf{x}^{\ast} = \textrm{arg min}_{\mathbf{x} \in \{0,1\}^N} \sum_{i \leq j} Q_{i,j}x_ix_j
\end{equation}
\hspace{35mm} for $Q_{i,j} \in \mathbb{R}$

We encode the QUBO formulation using a graph called logical graph. In the logical graph, each node corresponds to a binary variable $x_i$. We draw an edge between the nodes corresponding to two variables $x_i$ and $x_j$ if the quadratic coefficient $Q_{i,j}$ is nonzero.

The hardware graph is a representation of the topology of QPU with nodes corresponding to qubits and edges corresponding to qubits' couplers. QA system of D-Wave consists of three QPU topologies, namely Chimera - the earliest topology, Pegasus - the latest topology, and Zephyr - next generation QPU topology. These topologies are all in form of a grid of identical sets of nodes called unit cells. In this paper, we consider the Chimera topology. The reason is that \cite{Boothby2020} embedding methods in Chimera topology may be translated to Pegasus topology without modification, because Chimera is a subgraph of Pegasus. In other words, the method we develop for Chimera model also works for Pegasus topology. The Chimera topology is in form of $n\times m$ grid of $K_{c,c}$ unit cells as $T(n,m,c)$ .

QUBO models given optimization problems as a logical graph, and the topology of QPU used in QA as a hardware graph. In order to solve the final Hamiltonian which QA processes, we need to find an mapping from the logical graph to the hardware graph. Below we formally define the minor embedding problem:
\begin{definition}
(Minor embedding). Given a logical graph $P = (V_P, E_P)$ and a hardware graph $H = (V_H, E_H, T)$, minor embedding problem seeks to a mapping $\phi: V_P \rightarrow \mathcal{P}(V_H)$ satisfying three conditions:
\label{def:def1}
\end{definition}
\begin{enumerate}[leftmargin=*]
    \item \textbf{Chain connection}: We denote a subset of nodes in $H$ that are mapped from a node $u$ in $P$ as chain $\phi(u)$. Any subgraph $H'$ of $H$ induced by a chain $\phi(u)$ from H is a connected component with $\forall u \in V_P$.
    \item \textbf{Global connection}: For every edge $(u,v) \in E_P$, there exists at least one edge $(u', v') \in E_H$ such that $u' \in \phi(u)$, and $v' \in \phi(v)$.
    \item \textbf{One-to-many}: Two chains $\phi(u)$ and $\phi(v)$ in the hardware graph do not have any common nodes with $\forall u,v \in V_P, u \neq v$.
\end{enumerate}

Because hardware graphs based on QPU topologies are fixed-structured and incomplete graphs, finding minor embedding is a difficult problem.

\section{Adaptive topology embedding (ATOM)}\label{sec:method}

\ignoreme{

\mt{Change the section title, instead of using method. Let's go with the alg name as the title of this section. TAP: Toplogy expAnsion emBedding. or DATE: aDAptive Toplogy Embedding. or ATOM	Adaptive TOplogy eMbedding. Will think more...}

\mt{I have a very hard time understanding this alg even I already have knowledge about this alg. In 5 sentences, you can summary your algo, correct? In the overview section: As an abstract level, ATOM (whatever the name we chose) has xyz steps, name them. The goal of each, and how it work. Done!}

\mt{Then each subsection represents each subfunction. Here you have two sub-functions, embedding and expanding. Perhaps we need to have a better name}
\mt{Let's call Node Embedding for the first sub-function and Topology Adapting for the second sub-functions. Let's not call turns, that is simply iteration and no need to define. At the alg description, pseud-code and key points are enough. You don't need to  say $i$ to keep track the index, $w$ to store the index etc. Only point out the things you need for peopole to understand. THen when they look at the pseudo-code, they will know how to implement it if needed.}

\mt{Let's not call dead-end... Something else...}

\hoang{Actually, I want to define "turn" because if dead-end happens, one turn of embedding takes 2 iterations to finish (one to expanding and one to embedding.}

}

Here, we describe our algorithm, called Adaptice TOpology eMbedding (ATOM) for the minor embedding problem. We introduce the notion of adaptive topology of the hardware graph. We leverage the \emph{Chimera} topology to expand a hardware graph $T(n,m,c)$ to a new hardware graph $T(n',m',c)$ with $n' \geq n$, and $m' \geq m$ such that properties of an embedding in $T(n,m,c)$ is still preserved in $T(n',m',c)$. With this finding, we define adaptive topology as a subgraph of the hardware graph which grows along with size of current embedding. To be consistent with the Definition \ref{def:def1}, in which hardware graph $H$ is fixed, in our representation below, we assume that the hardware graph $H$ is large enough for every embedding.

\subsection{ATOM Algorithm}
Algorithm \ref{alg:alg1} outlines the ATOM algorithm for minor embedding. The algorithm takes the logical graph $P = (V_P, E_P)$ which encodes the QUBO formulation, and the hardware graph $H = (V_H, E_H, T(n,m,c))$ where $T(n,m,c)$ encodes its initial adaptive topology as input. It finds an embedding $\phi: V_P \rightarrow \mathcal{P}(V_H)$ which satisfies the three conditions of minor embedding provided in Definition \ref{def:def1}. We call such embedding \emph{feasible embedding}. Recall that a \emph{chain} $\phi(v)$ with $v \in P$ indicates the set of node in the hardware graph H embedded by $v$. Let us denote a subset of nodes in $P$ with $S$. We define $P[S]$ as the subgraph of $P$ induced on $S$. To be simple, we also assume that the logical graph P is a connected component.

The main idea behind ATOM is that nodes in $P$ are progressively embedded to $H$  at a time. We denote the process to embed one node to $H$ as \emph{turn} of embedding. We denote the embedded set after the $i$th turn with $S_i$, and the embedding after the $i$th turn with $\phi_i$. ATOM has three major steps: Initialization, Node Embedding, and Topology Adapting. Initialization determines the value of the set $S_0$, the embedding $\phi_0$, and other variables (lines 1 - 7). By finding a feasible embedding $\psi$ of the k-node subgraph $P'$ of $P$ with the most edges in $H$, ATOM initializes $S_0$ as $V(P')$, and $\phi_0$ as $\psi$. After initialization, at each turn, a new unembedded node is selected, and current embedding is updated by adding chains of the new node, and its neighbors in $H$ through \emph{Node Embedding} (lines 9 - 11). If the new node is successfully embedded to $H$, algorithm \ref{alg:alg1} updates current embedding with an additional embedding which includes new chains found in \emph{Node Embedding} (lines 16 - 20). Otherwise, algorithm \ref{alg:alg1} expands current embedding, and current adaptive topology through \emph{Topology Adapting}, and re-embeds the new node with expanded topology (lines 12 - 14). We explain the process of embedding a new node to $H$ in section B, and then describe the process of expanding current embedding in subsection C below. {\em In the rest of this paper, we omit the proofs of all lemmas and theorems due to page limit.}

\subsection{Node Embedding}
Algorithm \ref{alg:alg2} outlines how an additional embedding is found from current embedding at each turn. The algorithm takes the logical graph $P = (V_P, E_P)$, the hardware graph $H = (V_H, E_H, T(n,m,c))$, the embedding at current turn $\phi$, the embedded set at current turn $S$, and the new node $\Bar{v}$ to be included in the embedding as input. The goal of the algorithm is to find a new embedding $\phi'$ such that combination of $\phi'$ and $\phi$ is a feasible embedding for $P[S\cup \{\Bar{v}\}]$. Before going any further, we define the reverse mapping from a node in $H$ to a node in $P$ as $\phi^{-1}$. If a node $u \in H$ is not embedded by any node $v \in P$ through the embedding $\phi$, we call node $u$ as a \emph{free node} with $\phi^{-1}(u) = -1$. In addition, we define a path from $u \in V_{H}$ to the chain $\phi(v)$ with $v \in V_P$ with length of $L_{uv}$ as $Z_{uv} = \{z_0, z_1, z_2,\dots, z_{L_{uv}}\}$. If $Z_{uv}$ satisfies the condition in which for $0 \leq i \leq L_{uv} - 1$, $\phi^{-1}(z_i) = -1$ and, $\phi^{-1}(z_{L_{uv}}) = v$, the path $Z_{uv}$ is a \emph{clean path}. Finally, we define the subgraph of $H$ induced from adaptive topology $T(n,m,c)$ as $H'$.

Algorithm \ref{alg:alg2} splits the whole process into two major steps: (1) finding center for new chain of the new node $\Bar{v}$ (lines 1 - 8), and (2) assigning free node to connect the new chain to the existing embedding (lines 9 - 14). For the first step, after determining the embedded neighbour set of $\Bar{v}$ as $A = \{v|v \in S, (\Bar{v}, v) \in E_P\}$, algorithm \ref{alg:alg2} generates all possible clean shortest paths $Z_{uv}$ from a free node $u \in V_{H'}$ to chains $\phi(v)$ with $v \in A$ included in the set $Z$ by Breath First Search algorithm (lines 3 - 4). If from existing clean paths in $Z$, no feasible center can be found for the new chain (line 6), algorithm \ref{alg:alg2} returns an embedding $\phi'$ with $\phi'(v) = \emptyset$ for $\forall v \in P$, called \emph{empty embedding}. In contrast, the center of the chain for new node $\phi(\Bar{v})$ is selected as the free node $\Bar{u}$ to which sum of shortest clean paths from $\phi(v)$ with $v \in A$ is minimum (lines 7 - 8). After finishing the first step, we switch to the second step in which algorithm \ref{alg:alg2} assigns free nodes in clean paths from $\Bar{u}$ to existing chains of nodes in $P$, and includes these assignment in the additional embedding $\phi'$ (lines 9 - 13). Our node assignment does not only ensure to generate new global connections (Definition \ref{def:def1}) from the new chain $\phi(\Bar{v})$ to the existing embedding, and preserve chain connections (Definition \ref{def:def1}) in every chains, but it also distributes fairly the cardinality of new embedded nodes to the new chain, and existing chains based on degrees of nodes in $P$. As a result, it reduces the number of nodes needed in future embedding.

\begin{lemma}
Assume that $\phi_{i-1}$ is a feasible embedding for $P[S_{i-1}]$ and $\Bar{v}$ is new node to be embedded in the $i$th \emph{turn}, the embedding $\phi_{i}$ generated with $\phi_i(v) := \phi_{i-1}(v) \cup \phi'(v)$ for $\forall v \in P$ is a feasible embedding of $P[S_{i}]$ with $S_{i} = S_{i-1} \cup \{ \Bar{v} \}$ if embedding $\phi'$ is not \emph{empty embedding}.
\label{lem:lem1}
\end{lemma}

In some turns, it is possible that there exists no free node $\Bar{u} \in H$ such that all chains $\phi(v)$ with $v \in A$ can connect to $\Bar{u}$ by clean paths. We call it \textit{isolated problem}.

\begin{figure}[t]
    \includegraphics[width=1\columnwidth]{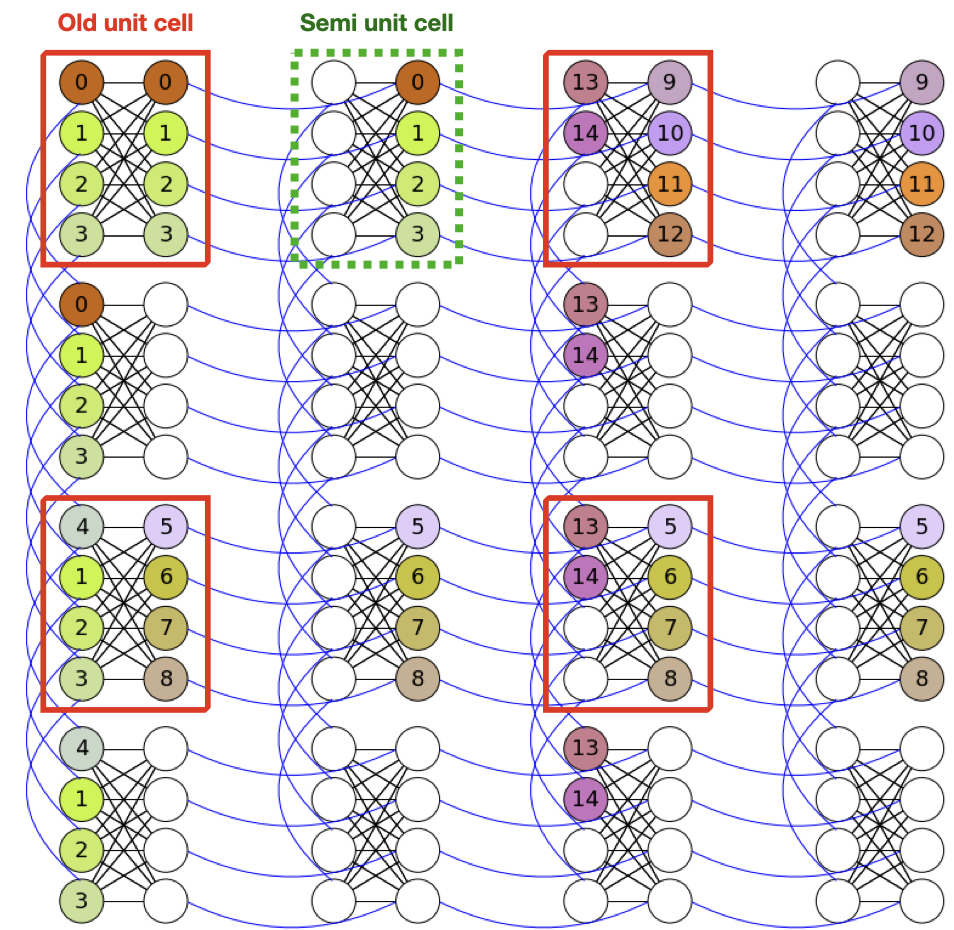}
    \caption{A hardware graph in form of $T(4,4,4)$ after expanding. Unit cells with red line border are old unit cells forming the hardware graph before expanding $T(2,2,4)$. The unit cell with green dash border is an example for semi unit cell used to connect old unit cells after expanding.}
    \label{fig:mesh1}
\end{figure}

\subsection{Topology Adapting}
Next, we show how to expand an adaptive topology to solve the \emph{isolated problem}. Recall that using the Chimera model, the adaptive topology of $H$ as $T(n,m,c)$ is an $n \times m$ grid of $K_{c,c}$ unit cells. Each unit cell $K_{c,c}$ is a bipartite graph with c nodes on each partite. Each node $u \in H$ has a coordinate $(x_u, y_u, z_u)$ in the topology $T(n,m,c)$ with $x_u \in [0, n-1]$ indicates the row index, $y_u \in [0, m-1]$ indicates the column index, and $z_u \in [0, 2c-1]$ indicates the bipartite index. In a unit cell, nodes on the right partite have connections to corresponding nodes in two adjacent unit cells in the same row while nodes on the left partite have connections to corresponding nodes in two adjacent unit cells in the same column (see in Figure \ref{fig:mesh1}). 

Algorithm \ref{alg:alg3} presents the topology adapting algorithm. The algorithm takes current embedding $\phi$, and current adaptive topology $T(n,m,c)$ as input. The current embedding of nodes located in the unit cell $(x,y)$ in $H$ is shifted by $x$ rows and $y$ columns to the bottom right corner of $H$ (lines 4 - 6). As a result, there are two kinds of unit cells after expanding, called \emph{empty unit cells} whose nodes are not assigned to any chains and \emph{old unit cells} whose nodes are assigned to chains from the embedding $\phi$. Algorithm \ref{alg:alg3} then, assigns nodes in empty unit cells located in between two old unit cells. These unit cells are called \emph{semi unit cells}. Then, we denote semi unit cells in between two old unit cells in the same row as \emph{row semi unit cells}, and semi unit cells in between two old unit cells in the same column as \emph{column semi unit cells}. The rule of assignment is that nodes on the right partite of a row semi unit cell $(x,y)$ are assigned to the same chain with the corresponding node in the adjacent left old unit cell $(x-1,y)$ while nodes on the left partite of a column semi unit cell $(x,y)$ are assigned to the same chain with the corresponding node in the adjacent top old unit cell $(x,y-1)$. Figure \ref{fig:mesh1} shows the expanding process. We realize that semi unit cells are bridges to keep chain connections, and global connections from $\phi$ after expanding. It implies that the embedding $\phi^\dagger$ expanded from $\phi_i$ is a feasible embedding of $P[S_i]$.

\begin{lemma}
After expanding the current embedding $\phi$ to new embedding $\phi^\dagger$ by the Algorithm \ref{alg:alg3}, the addition embedding $\phi'$ found by the Algorithm \ref{alg:alg2} with current embedding as $\phi^\dagger$ is not \emph{empty embedding}.
\label{lem:lem2}
\end{lemma} 

We conclude from Lemma \ref{lem:lem2} that our method avoids \textit{isolated problem} after expanding the current embedding.
\begin{theorem}
Given $S_i$ as the embedded set, and $\phi_i$ as the embedding found by the Algorithm \ref{alg:alg2} after the $i$th turn, $\phi_i$ is a feasible embedding for $P[S_i]$.
\label{theo:theo1}
\end{theorem} 

\ignoreme{
\begin{proof}
From Lemma \ref{lem:lem1} \& \ref{lem:lem2}, and the fact that Theorem \ref{theo:theo1} holds for $i = 0$, by induction, we conclude that Theorem \ref{theo:theo1} holds for $i > 0$. 
\end{proof}
}
Theorem \ref{theo:theo1} implies that the resulting embedding $\phi$ after $|V_P| - |V(P')|$ turns is a feasible embedding of $P$ in $H$.

\begin{theorem}
The Algorithm \ref{alg:alg1} returns a feasible embedding for $P$ after at most $3|V_P|$ iterations.
\label{theo:theo2}
\end{theorem}
\ignoreme{
\begin{proof}
Lemma \ref{lem:lem2} implies that each turn of embedding costs at most three iterations, so Algorithm \ref{alg:alg1} returns a feasible embedding after at most $3(|V_P| - |V(P')|) \leq 3|V_P|$ iterations.
\end{proof}
}

\begin{algorithm}
\DontPrintSemicolon
  \KwInput{Logical graph $P = (V_P, E_P)$, hardware graph $H = (V_H, E_H, T(n,m,c))$.}
  \KwOutput{A feasible embedding $\phi: V_P \rightarrow \mathcal{P}(V_{H})$}
  
    Let the subgraph of $H$ induced by $T(n,m,c)$ as $H'$.

    Let the densest k-subgraph of $P$ be $P'$.
    
    Let the complete embedding of $P'$ in $H'$ be $\psi$.
    
    Initialize $\phi_0$ such that $\phi_0(v) := \psi(v)$ for $\forall v \in V_P$.
    
    $S_0 := V(P')$
    
    $i := 1$
    
    $w[v] := 0$ for $\forall v \in V_P$
    
    $isolated := False$
    
    \While{$ i \leq |V_P| - |V(P')|$}
    {
        \If{not $isolated$}{
            $\Bar{v}$ := $\textrm{arg min}_{v \in V_P \setminus S} \sum_{u \in S, (u,v) \in E_P} w_u$\newline 
        }
        $\phi'$ := NODE\_EMBEDDING$(P, H, \phi_{i-1}, S_{i-1}, \Bar{v})$
        
       \If{$\exists v \in V_P$ such that $\phi'(v) \neq \emptyset$}{
           $isolated$ := False
            
            Initialize $\phi_i$ such that $\phi_i(v) := \phi_{i-1}(v) \cup \phi'(v)$ for $\forall v \in V_P$
            
            $S_i := S_{i-1} \cup \{\Bar{v}\}$;
            $w[\Bar{v}] := i$; $i:= i+1$
       }
       \Else
       {
            $isolated$ := True
           
            $\phi_{i-1}, T$:= TOPOLOGY\_ADAPTING$(H, \phi_{i-1})$
        }
    }
    \Return $\phi_{|V_P| - |V(P')|}$
\caption{Algorithm ATOM}
\label{alg:alg1}
\end{algorithm}

\begin{algorithm}[!ht]
\DontPrintSemicolon
  \KwInput{Logical graph $P = (V_P, E_P)$, hardware graph $H = (V_H, E_H, T(n,m,c))$, current embedding $\phi$, current embedded set S, new node $\Bar{v}$}
  \KwOutput{Additional embedding $\phi'$}
  
  Let the subgraph of $H$ induced by $T(n,m,c)$ as $H' = (V_{H'}, E_{H'})$.
  
  Let the embedded neighbour set of $\Bar{v}$ be $A = \{v|v \in S, (\Bar{v}, v) \in E_P\}$.
  
  Let $Z_{uv} := \{z_0, z_1, z_2, ..., z_{L_{uv}}\}$ be the clean shortest path from node u to the chain $\phi(v)$.
  
  Let $Z$ be the set of all possible clean shortest paths $Z_{uv}$ with $u \in V_{H'}, v \in A$.
  
  Initialize $\phi'$ with $\phi'(v)$ := $\emptyset$ for $\forall v \in V_P$.
  
  \If{$\nexists u \in V_{H'}$ such that $\exists Z_{uv} \in Z$ for $\forall v \in A$}{
    \Return $\phi'$
  }
  $\Bar{u} := \textrm{argmin}_{u \in V_{H'}, \phi^{-1}(u) = -1} \sum_{v \in A} |Z_{uv}|$
  
  $\phi'(\Bar{v}) \gets \phi'(\Bar{v}) \cup \{\Bar{u}\}$
  
  \For {$v \in A$} {
    Find the biggest index $i_v$ from 1 to $L_{\Bar{u} v} - 1$, such that there exists $v' \neq v, v' \in A, z_i \in Z_{\Bar{v} v'}$.
    
    $\delta := \lfloor \frac{\gamma_{\Bar{v}}}{\gamma_{\Bar{v}} + \gamma_{v}}(L_{\Bar{u} v} - 1 - i) \rfloor$ 
    
    $\phi'(\Bar{v}) \gets \phi'(\Bar{v}) \cup \{z_i| 1 \leq i \leq i_v + \delta\}$
    
    $\phi'(v) \gets \phi'(v) \cup \{z_i| i_v + \delta + 1 \leq i \leq L_{\Bar{u} v}\}$
    }
    \Return $\phi'$
\caption{NODE EMBEDDING}
\label{alg:alg2}
\end{algorithm}

\begin{algorithm}[!ht]
\DontPrintSemicolon
  \KwInput{Hardware graph $H = (V_H, E_H, T(n,m,c))$, current embedding $\phi$}
  \KwOutput{An expanded embedding $\phi^\dagger: V_P \rightarrow \mathcal{P}(V_{H})$, and the adaptive topology T after expanding}

  Initialize $\phi^\dagger$ such that $\phi^\dagger(v) := \emptyset$ for $\forall v \in V_P$.
  
  \For{$v \in V_P$}{
    \For{$u \in \phi(v)$} {
    Let $(x,y,z)$ be the coordinate of $u$.
    
    Let $u' \in H$ with coordinate of $(2x, 2y, z)$.
    
    $\phi^\dagger(v) \gets \phi^\dagger(v) \cup \{u'\}$
    
    \If{$z \leq c - 1$}
    {
        Let $u'' \in H$ with coordinate $(2x + 1, 2y, z)$
        
        $\phi^\dagger(v) \gets \phi^\dagger(v) \cup \{u''\}$
    }
    \Else
    {
        Let $u'' \in H$ with coordinate $(2x, 2y+1, z)$
        
        $\phi^\dagger(v) \gets \phi^\dagger(v) \cup \{u''\}$
    }
    }
  }
  
  \Return $\phi^\dagger$, T(2n, 2m, c)

\caption{TOPOLOGY ADAPTING}
\label{alg:alg3}
\end{algorithm}

\section{Experiments}\label{sec:exp}

In this section, we evaluate our method on synthetic dataset. We compare our method to two state-of-the-art methods: OCT-based \cite{Goodrich2018} , and Minorminer \cite{Cai2014} which was developed by D-Wave. Minorminer is the best heuristic method as of now for sparse logical graphs.

We construct logical graphs using three models, namely, Barabási-Albert with initialization by star graphs ($BA_{star}$), Barabási-Albert with initialization by complete graphs ($BA_{complete}$), and d-Regular. For each model, we generate 38 logical graphs by varying the number of nodes, and average node degree $d$ as $|V_P| \in \{100, 200,\dots, 1900\}$ and degree $d \in \{10, 20\}$.
    \ignoreme{\item \textit{Real world instances}. We collect logical graphs encoding Max-Cut instances from MQLib\cite{DunningEtAl2018}. In more details, we select instances with the number of nodes $|V_P| \leq 2000$ including 15 instances from Culberson source\cite{Culberson1995} and 50 instances from Beasley source\cite{Beasley1990}.}

We use the following three criteria to evaluate our method, and the competing methods.

\begin{figure*}[t]
\begin{subfigure}{.32\textwidth}
  \centering
  \includegraphics[width=1\textwidth]{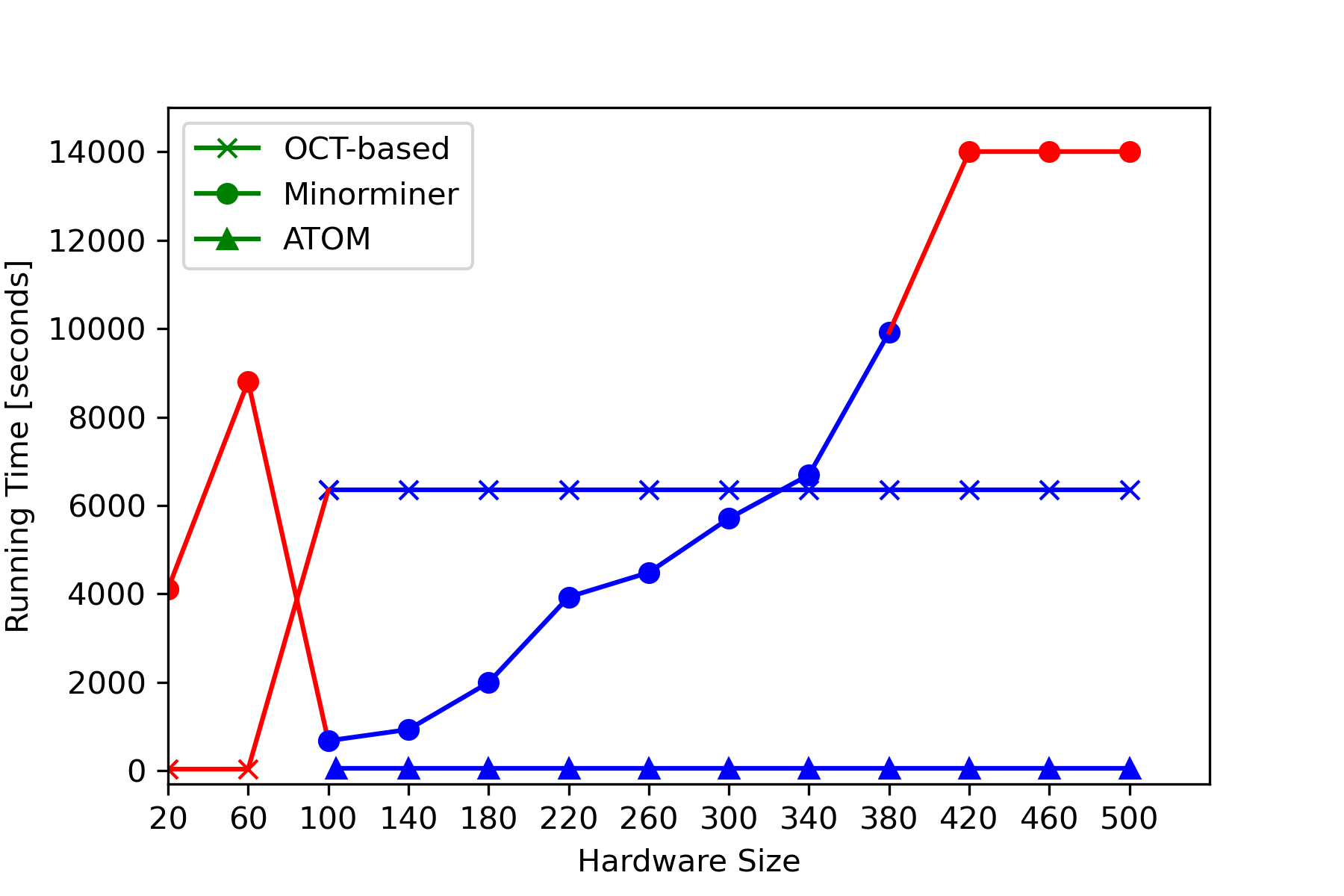}
  \caption{$\#$ of nodes = 400}
  \label{fig:sfig1}
\end{subfigure}
\hfill
\begin{subfigure}{.32\textwidth}
  \centering
  \includegraphics[width=1\columnwidth]{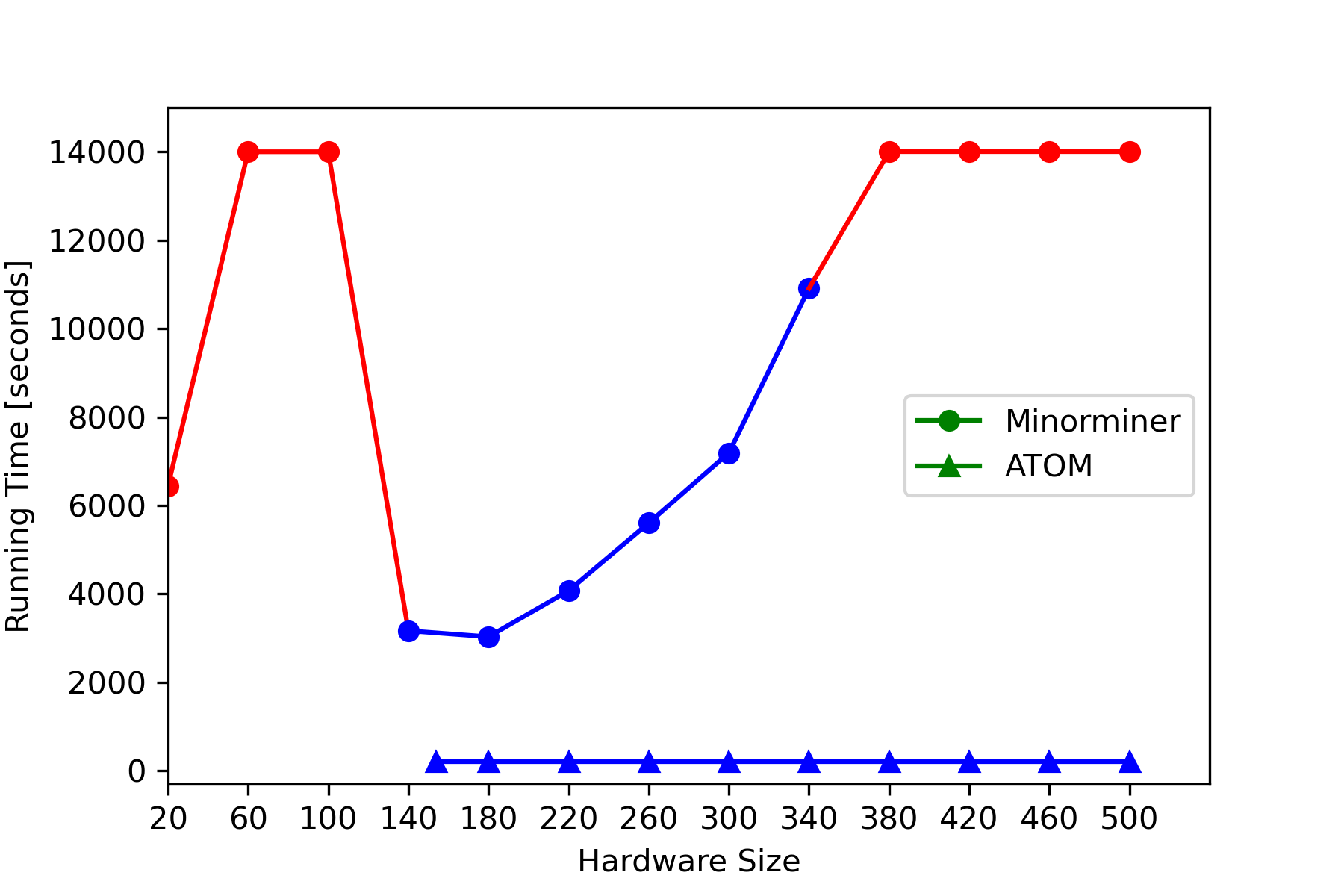}
  \caption{$\#$ of nodes = 600}
  \label{fig:sfig2}
\end{subfigure}
\hfill
\begin{subfigure}{.32\textwidth}
  \centering
  \includegraphics[width=1\columnwidth]{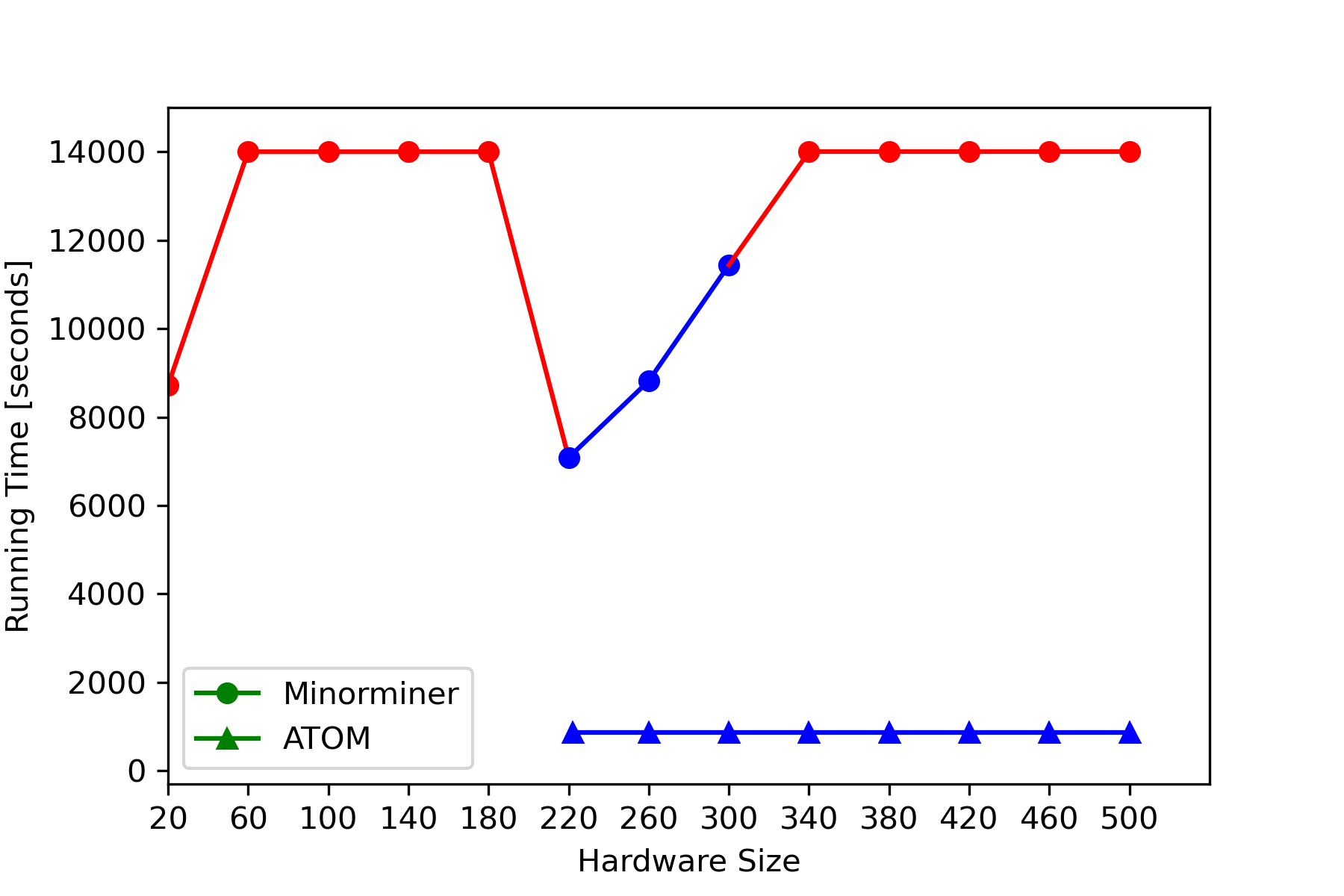}
  \caption{$\#$ of nodes = 900}
  \label{fig:sfig2}
\end{subfigure}
\caption{Running time of OCT-based, Minorminer, and ATOM for different hardware sizes. Logical graphs are generated using $BA_{complete}$ model with the number of nodes = 400, 600, 900 in \textbf{(a)}, \textbf{(b)}, \textbf{(c)} respectively. Blue points represents for hardware graph size for which the corresponding method can find a feasible embedding. Red points represents infeasible cases. In \textbf{(b)} and \textbf{(c)}, we do not report OCT-based as it cannot return any feasible embedding.}
\label{fig:fig2}
\end{figure*}

\begin{figure*}[t]
\begin{subfigure}{.32\textwidth}
  \centering
  \includegraphics[width=1\textwidth]{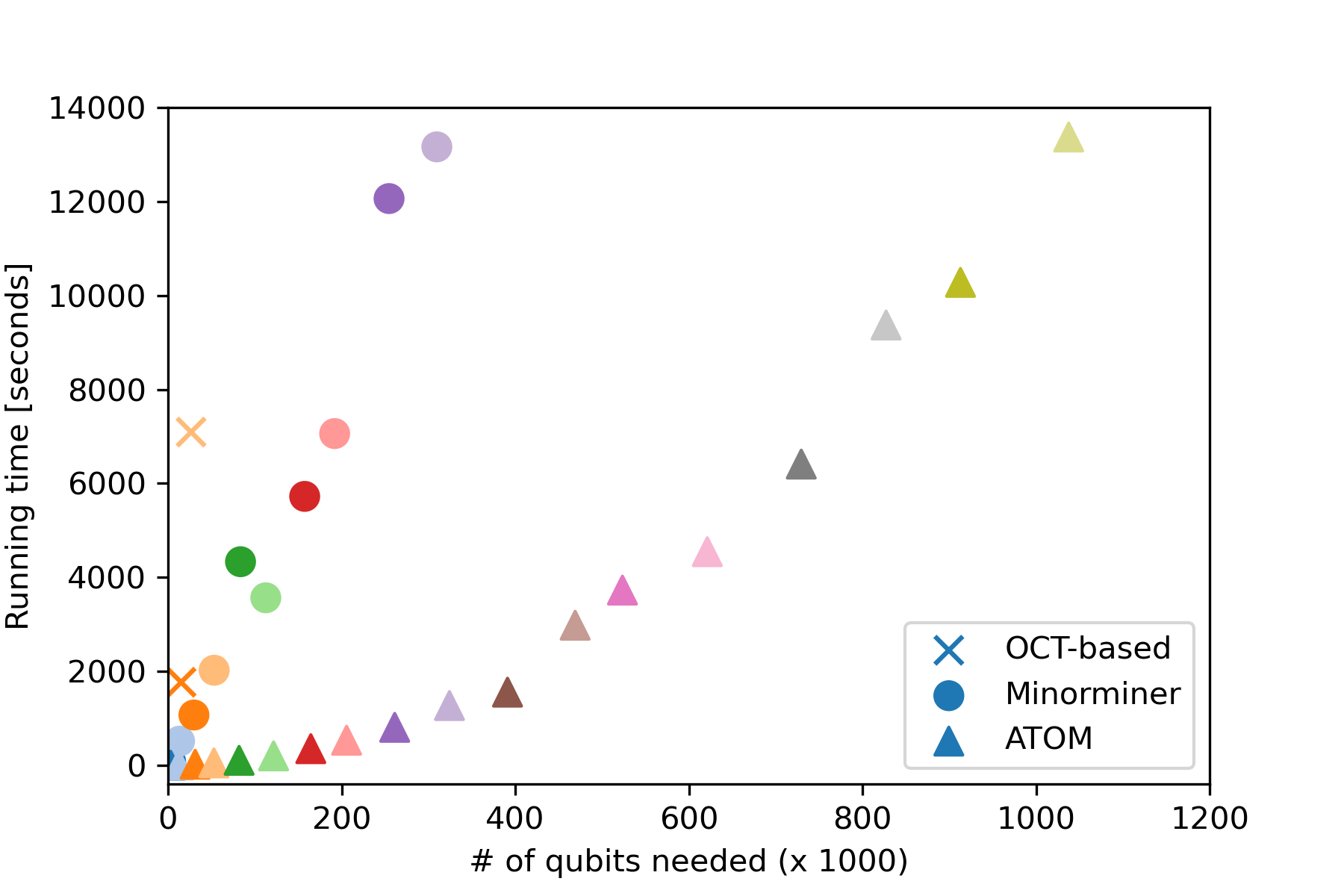}
  \caption{$BA_{star}$ with $d = 10$}
  \label{fig:sfig1}
\end{subfigure}
\hfill
\begin{subfigure}{.32\textwidth}
  \centering
  \includegraphics[width=1\columnwidth]{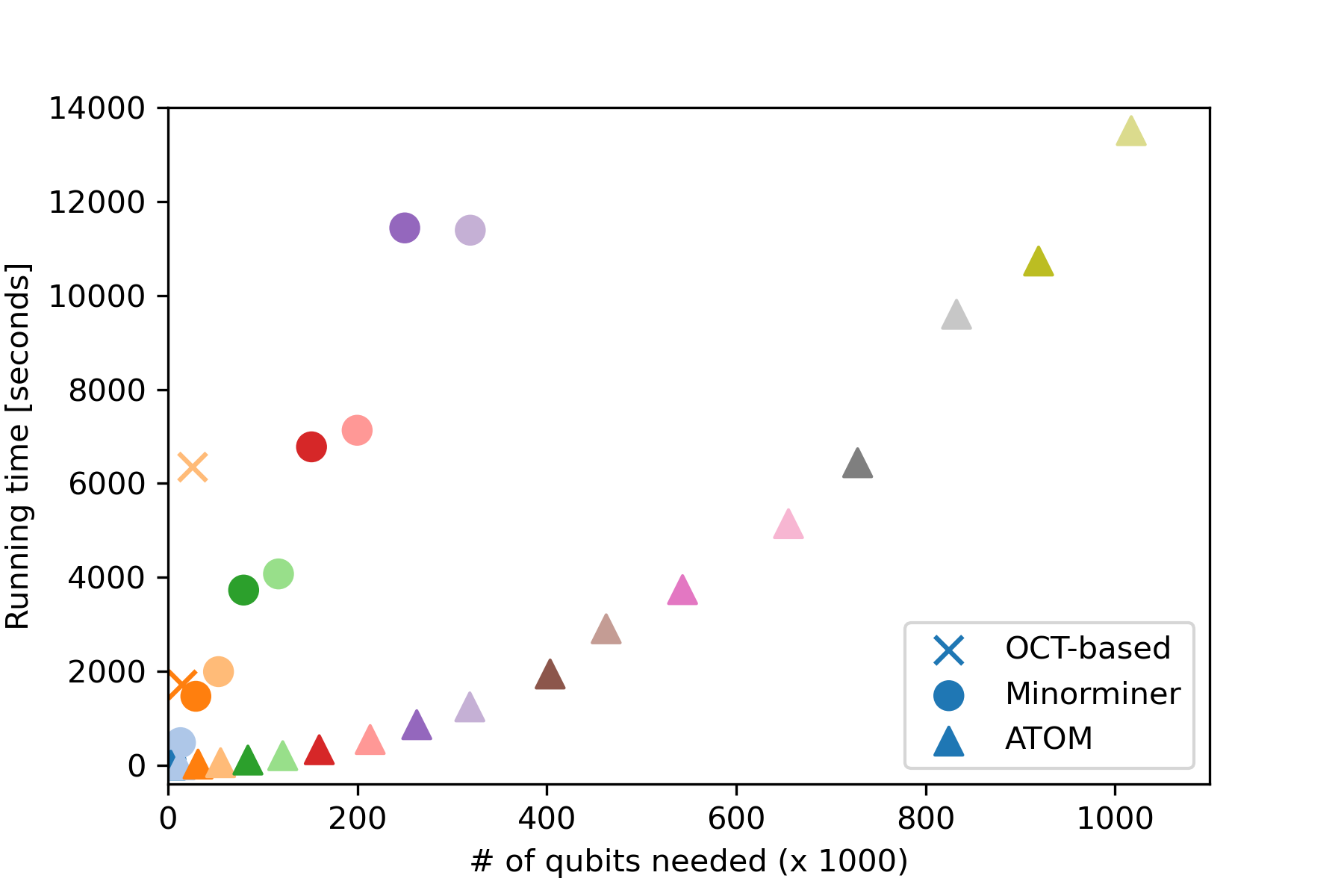}
  \caption{$BA_{complete}$ with $d = 10$}
  \label{fig:sfig2}
\end{subfigure}
\hfill
\begin{subfigure}{.32\textwidth}
  \centering
  \includegraphics[width=1\columnwidth]{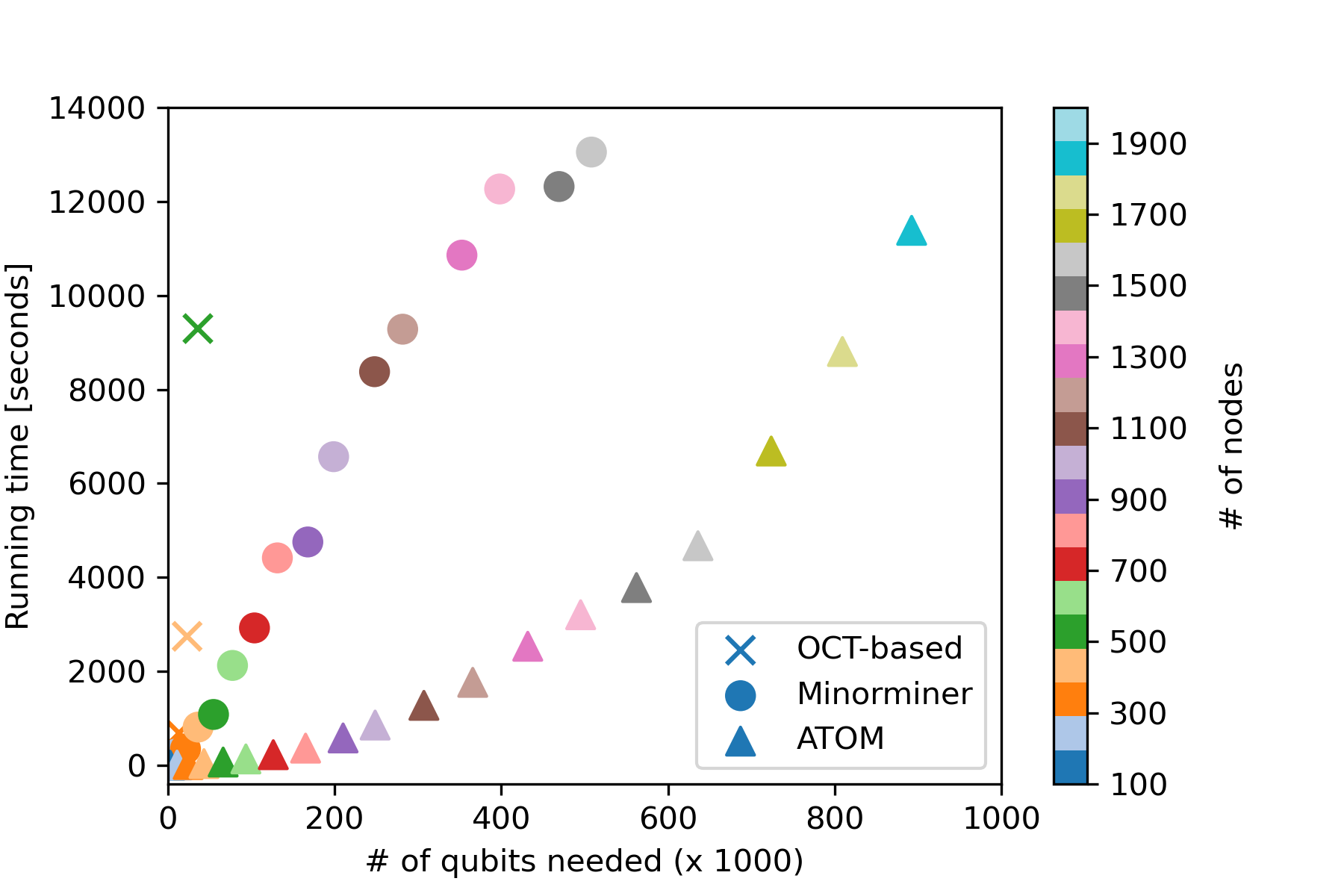}
  \caption{$Regular$ with $d = 10$}
  \label{fig:sfig3}
\end{subfigure}

\begin{subfigure}{.32\textwidth}
  \centering
  \includegraphics[width=1\textwidth]{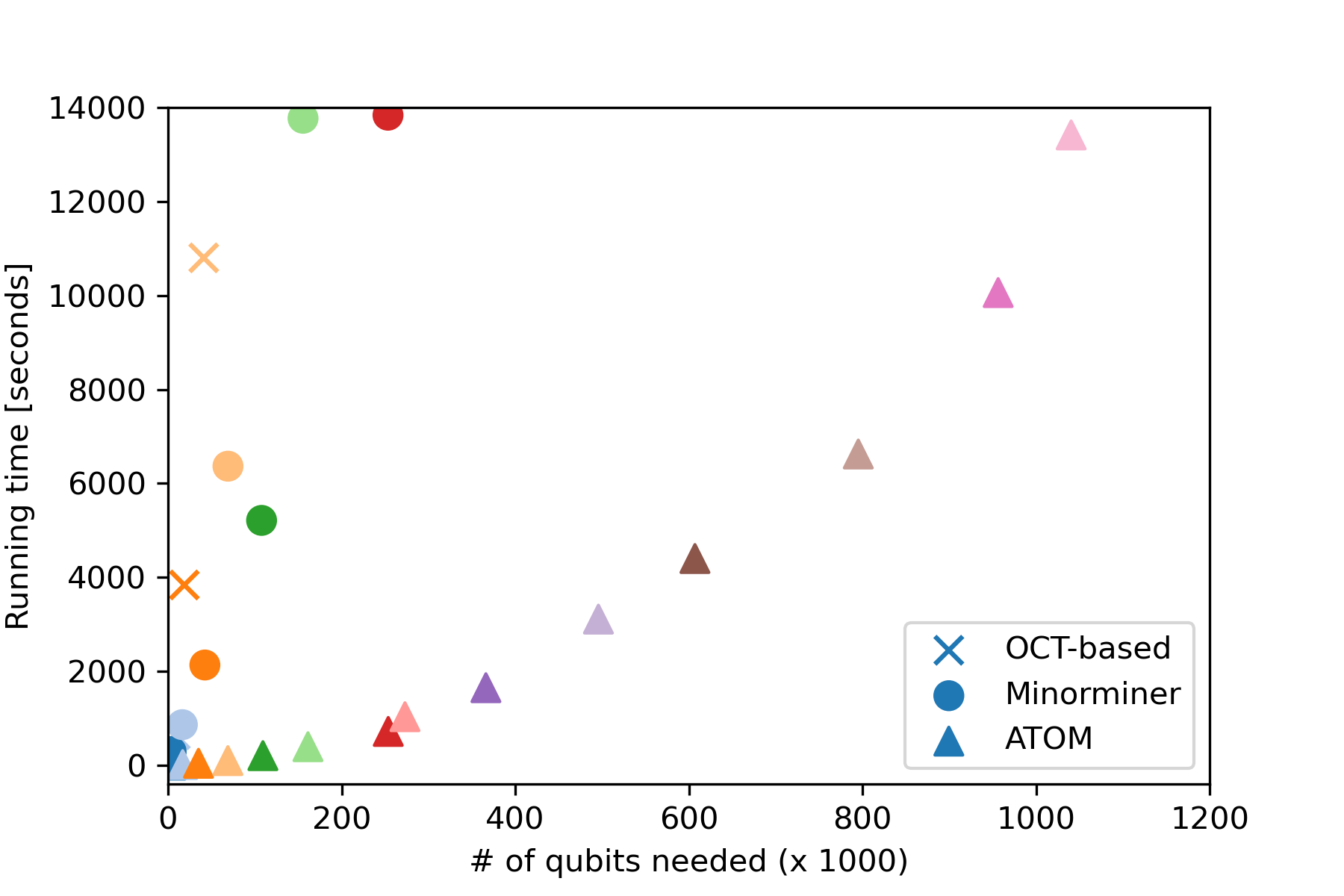}
  \caption{$BA_{star}$ with $d = 20$}
  \label{fig:sfig4}
\end{subfigure}
\hfill
\begin{subfigure}{.32\textwidth}
  \centering
  \includegraphics[width=1\columnwidth]{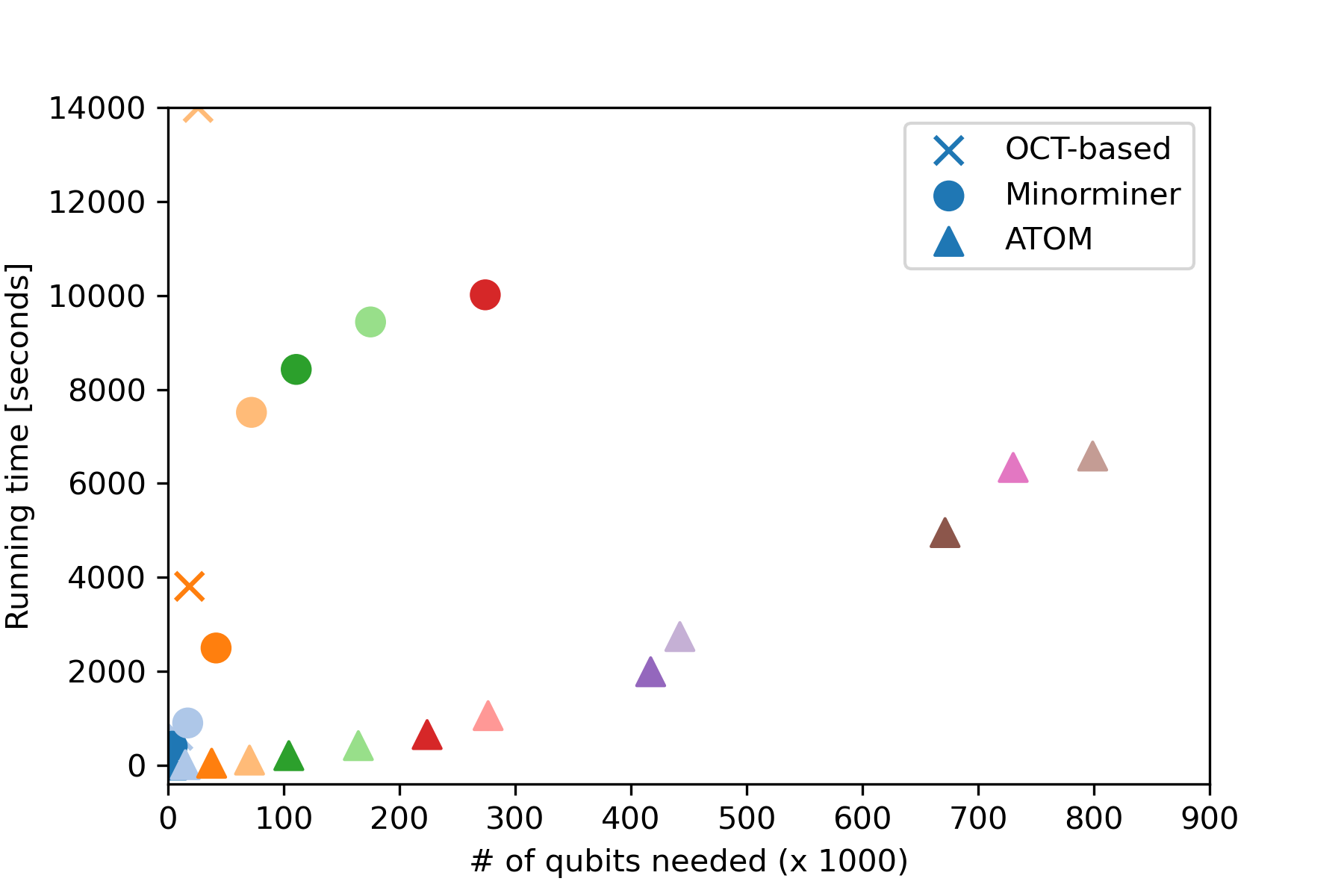}
  \caption{$BA_{complete}$ with $d = 20$}
  \label{fig:sfig5}
\end{subfigure}
\hfill
\begin{subfigure}{.32\textwidth}
  \centering
  \includegraphics[width=1\columnwidth]{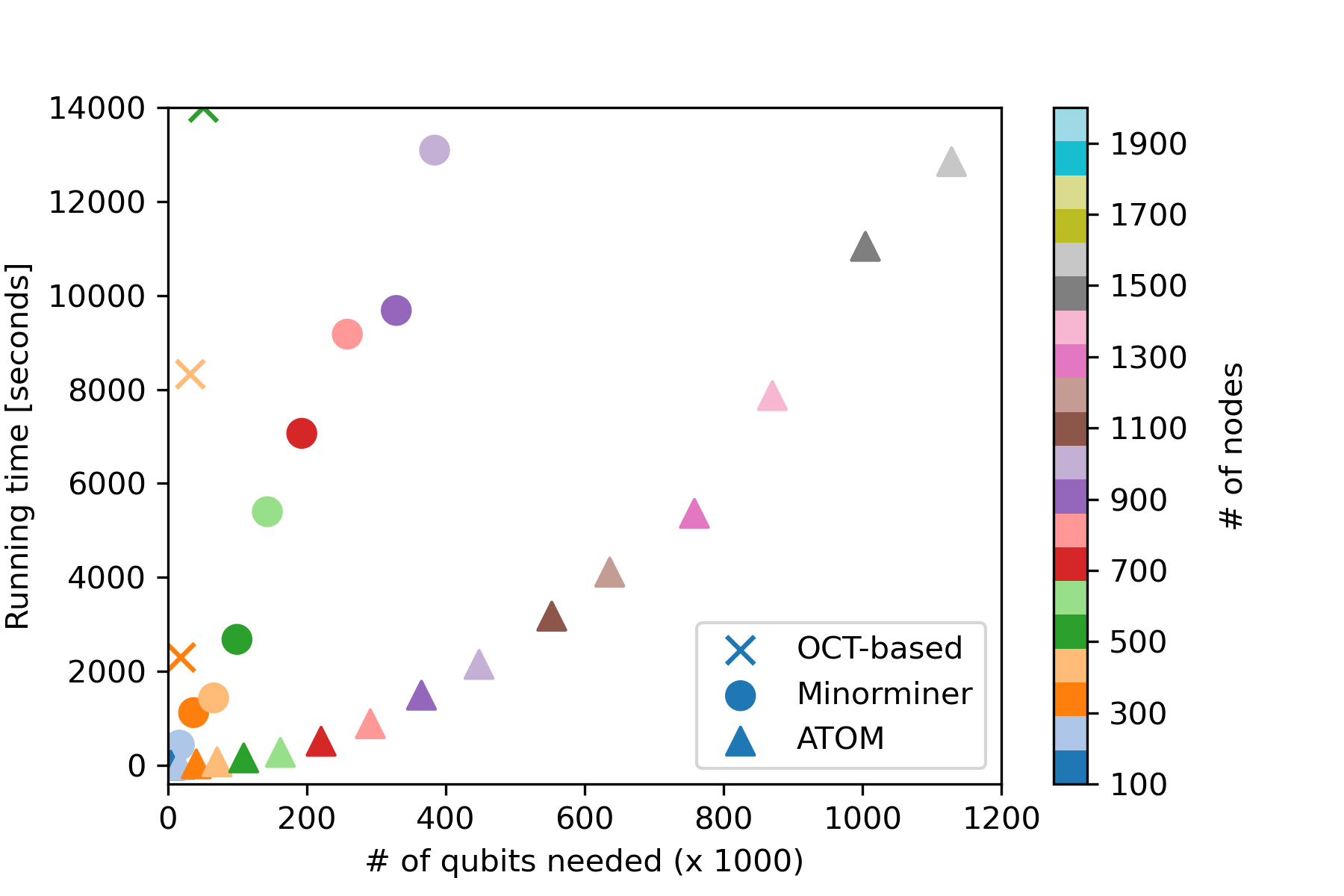}
  \caption{$Regular$ with $d = 20$}
  \label{fig:sfig6}
\end{subfigure}
\caption{Running time and the number of qubits needed for embedding synthetic instances using $BA_{star}$, $BA_{complete}$, and $Regular$ graph models for  degree $d \in \{10,20\}$. Different colors indicate sizes of logical graphs ranging from 100 to 1900.}
\label{fig:fig3}
\end{figure*}

\begin{enumerate}[leftmargin=*]
    \item \textit{Minimum hardware size}. We compute the smallest hardware size needed to include the resulting embedding. Since Minorminer, and OCT-based take a specific hardware size as input, we run their method with the hardware graph following the topology $T(n,n,4)$ with $n \in \{20, 60,\dots,500\}$. Unlike these methods, ATOM adapts the hardware size to the given problem instance. Thus, we report the size of final adaptive topology as the minimum hardware size of ATOM.  
    
    \item \textit{Running time}. We measure the time needed to return a feasible embedding, or the time needed to claim that no feasible embedding exists for the given hardware size. We set the \emph{time limit} to 14000 seconds.
    
    \item \textit{The number of qubits needed}. We measure the number of nodes in $H$ which are mapped by a node in $P$ in the resulting embedding. 
\end{enumerate}

\textbf{Environment.} We implement ATOM in C++. We use  Minorminer, and OCT-based code provided by their authors. For OCT-based method, we select \emph{Fast-OCT-reduce} version as it is the most effective version of OCT-based. We run all experiments on a Debian GNU/Linux machine with Intel(R) Xeon(R) CPU E5-2697 v4 @ 2.30GHz.


\noindent
\paragraph{Evaluation of the effect of hardware size}
Recall that the fundamental contribution of this paper is that our method adapts to the given hardware size unlike existing state-of-the-art methods. We first evaluate how this affects the cost of solving the minor embedding problem. 
Figure \ref{fig:fig2}, shows the minimum hardware size needed to find a feasible embedding from OCT-based, Minorminer, and ATOM for different logical and hardware graph sizes. Our resuts demonstrate that ATOM outperforms competing methods for all logical and hardware graph sizes. The performance of both OCT-based, and Minorminer depends on how we select hardware sizes, while ATOM computes the minimum hardware size from final embedding. We observe that, OCT-based, and Minorminer return infeasible solutions if selected hardware size is too small. In addition, Minorminer can result in large running time for inappropriate hardware sizes. Second, minimum hardware size from ATOM is comparable to two other methods in all cases. Thus, we conclude that ATOM is effective for finding embedding on real QPU topology whose size is always fixed.

\noindent
\paragraph{Evaluation of the number of qubits needed}
The success of a QA algorithm depends on the number of qubits it uses as well as the time it takes to embed the given optimization problem to so many qubits. Ideally, it is preferable to embed into a small number of qubits in short amount of time. 
In figure \ref{fig:fig3}, we show the relation between the running time, and the number of qubits needed for three methods. We observe that ATOM outperforms two others in running time for all cases.
On average, ATOM is up to 20 times faster than Minorminer, and up to 66 times faster than OCT-based. 

As a result, with the timeout is 14000 seconds, ATOM is able to find feasible embeddings for logical graphs with size of 1800, and 1500 for $d = 10$, and $d = 20$ respectively, while the corresponding numbers for Minorminer is 1500, and 1000, and for OCT-based is 500, and 400. The number of qubits needed for embedding resulted from ATOM is comparable with OCT-based, and Minorminer, and even better in some cases. Therefore, ATOM is able to provide a feasible embedding for QA much faster than Minorminer, and OCT-based can, but do not require much more additional resources. That leads to more economic advantages since with ATOM, people can solve more optimization problems in QA at the same time.


\ignoreme{
\begin{figure}
\begin{subfigure}{.24\textwidth}
  \centering
  \includegraphics[width=1\textwidth]{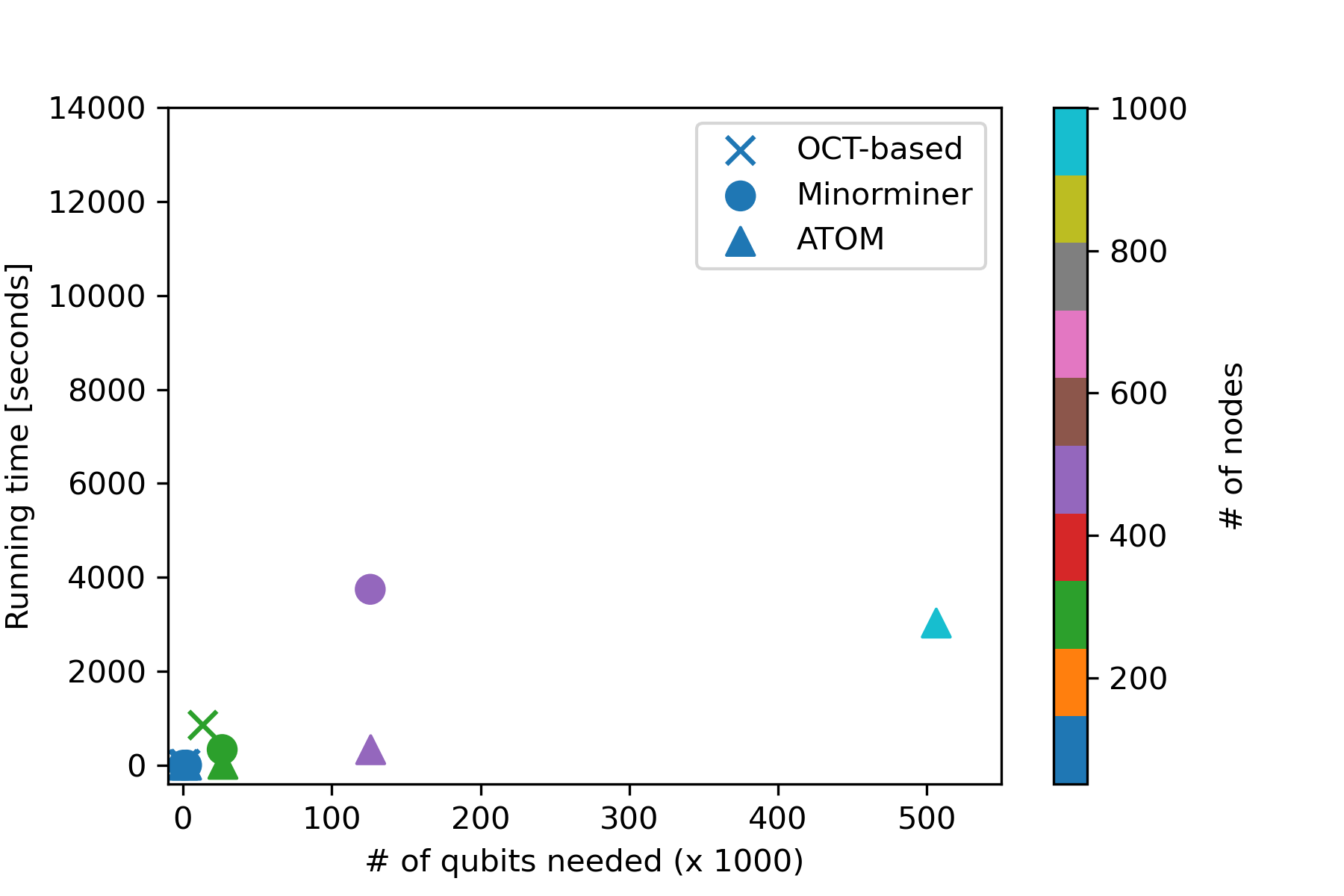}
  \caption{$Beasley$}
  \label{fig:sfig1}
\end{subfigure}
\hfill
\begin{subfigure}{.24\textwidth}
  \centering
  \includegraphics[width=1\columnwidth]{}
  \caption{$Culberson$}
  \label{fig:sfig2}
\end{subfigure}
\caption{Comparisons of running time and the number of qubits needed for embedding synthetic instances with two different sources $Beasley$, and $Culberson$ corresponding to \textbf{(a)}, and \textbf{(b)}. Different colors are to indicate sizes of logical graphs ranging from 100 to 1900.}
\label{fig:fig}
\end{figure}
}

\section{Conclusion}\label{sec:con}

In this work, we introduce a novel notion of adaptive topology, and propose a minor embedding algorithm, namely Adaptive TOpology eMbedding (ATOM). ATOM is able to provide a feasible embedding without fixing the hardware graph size. Our experimental results show that ATOM requires much less running time to find a feasible embedding without demanding additional resources, compared with two state-of-the-art methods: OCT-based, and Minorminer. In the future, along with the rapid expansion in QPU size, and increasing demand on solving large optimization problems quickly, ATOM will be an efficient choice for minor embedding in QA.




\bibliographystyle{ieeetr}
\bibliography{bibliography.bib}
\end{document}